\documentclass[%
reprint,
superscriptaddress,
amsmath,amssymb,amsfonts,
aps,
prl,
floatfix,notitlepage,latexsym
]{revtex4-1}

\usepackage[dvipdfmx]{graphicx}% Include figure files
\usepackage{dcolumn}% Align table columns on decimal point
\usepackage{bm}% bold mathematica
\usepackage{braket}
\usepackage{mediabb} 
\usepackage[dvipdfmx]{attachfile2}
\begin{abstract} 
Non-collinear chiral antiferromagnets like Mn${}_{3}$Sn and Mn${}_{3}$Ge are known to show gigantic anomalous Hall response depending on the orientation of their inverse chiral magnetic order of Mn atoms in Kagom\'{e} layers. Here we study the stability of such magnetic order in the absence of external magnetic fields on the basis of stochastic Landau-Lifshitz-Gilbert equation for a simplified two-dimensional model of these materials. 
We find that even without external magnetic fields, the ordered state is, once formed, highly stable against thermal fluctuations. Moreover, we show that if Mn spins are well confined inside each Kagom\'{e} layers, by injecting spin-current using spin-filtering effect of ferromagnetic metals, we can control the in-plane magnetic structure in a field free way. 

\vspace{3mm}
\noindent PhySH: Antiferromagnetism, Magnetization switching, Spintronics, LLG
\end{abstract}

\begin{document}

%%%%%%%%%%%%%%%%%%%%%%%%%%%%%%%%%%%%%%%%%%%%%%%%%
% Paper Information
%%%%%%%%%%%%%%%%%%%%%%%%%%%%%%%%%%%%%%%%%%%%%%%%%

\title{Field-free, spin-current control of magnetization \\ in non-collinear chiral antiferromagnets}
\author{Hiroyuki Fujita}
\affiliation{Institute for Solid State Physics, University of Tokyo, Kashiwa 277-8581, Japan}
\email{h-fujita@issp.u-tokyo.ac.jp}
\date{\today}

\maketitle
\begin{figure}[htbp]
   \centering
\includegraphics[width = 65mm]{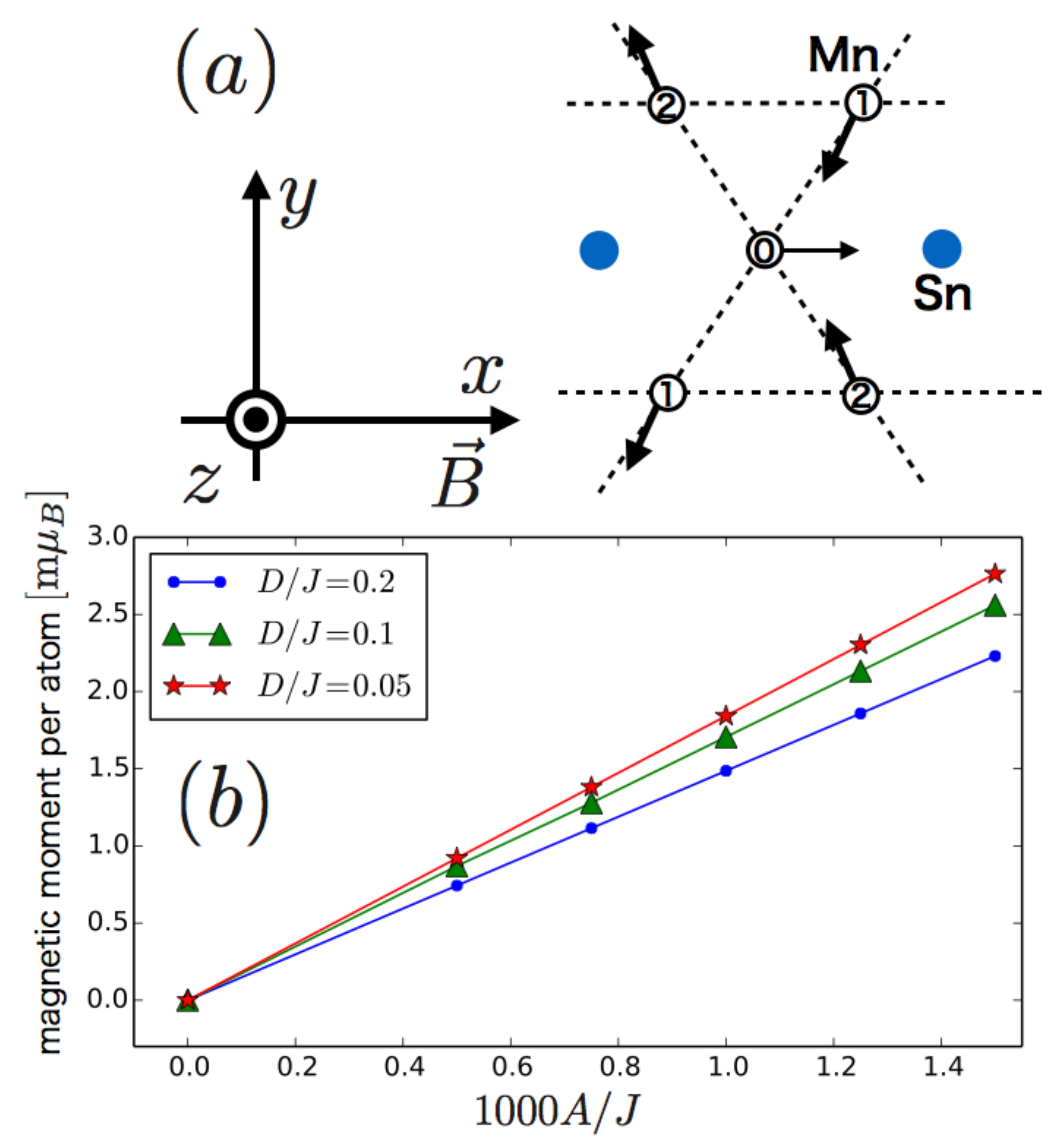}
%       \caption{Equilibrium spin configuration under the applied external magnetic field (in the $x$-direction). Black filled circles represent Mn atoms and blue filled circles do nonmagnetic atoms (Sn atoms for Mn${}_{3}$Sn).}
       \caption{Spin structure of the model \eqref{Hamiltonian} at zero temperature. ($a$) Ground state spin configuration under the applied external magnetic field in the $x$-direction. ($b$) Numerically obtained weak ferromagnetic moment per atom in the ground state without the external field as a function of the anisotropy energy $A$.}
          \label{spin config} 
  \end{figure}

{\it Introduction ---}
Antiferromagnetic Mn compounds like Mn${}_{3}$Sn, Mn${}_{3}$Ge have a structure of stacked alternate Kagom\'{e} layers of Mn atoms~\cite{doi:10.1143/JPSJ.51.2478,Tomiyoshi:1982aa,Nagamiya:1982aa,Sandratskii:1998aa}. In these materials, at moderate temperature localized Mn spins are confined inside Kagom\'{e} layers and develop inverse chiral order [see Fig.~\ref{spin config}($a$)]. The inverse chiral order is stabilized by antiferromagnetic exchange interaction and Dzyaloshinskii-Moriya (DM) interaction~\cite{DZYALOSHINSKY1958241,PhysRev.120.91}. Due to the in-plane anisotropy in the direction of the nearest nonmagnetic atoms for each Mn atom, in spite of the antiferromagnetic nature, net magnetization of each triangle unit cell of Kagom\'{e} layers does not completely cancels out. As a result, the system shows weak ferromagnetism ($\sim 2$ m$\mu_{B}$ per Mn atom for Mn${}_{3}$Sn~\cite{Nakatsuji:2015aa}).
 
 The non-collinear antiferromagnetic phase of those materials gathers growing attentions in the context of antiferromagnetic spintronics~\cite{Jungwirth:2016aa,Baltz2016,Wadleyaab1031} because they show gigantic Hall responses even at room temperature~\cite{Nakatsuji:2015aa,PhysRevApplied.5.064009,Nayake1501870} when in-plane external magnetic fields are applied. These experimental observations are consistent with recent theoretical works~\cite{PhysRevLett.112.017205,refId0} and imply Weyl physics~\cite{Yang2016}. The transport properties of those materials are closely related to the magnetic structure. For example, as is experimentally demonstrated, by reversing the direction of the applied external magnetic field, we can switch the sign of Hall signal. Due to the extremely small anisotropy energy of the inverse chiral structure~\cite{Nakatsuji:2015aa}, even a weak magnetic field can rotate the in-plane magnetic structure and that offers a way of controlling transport properties. 
 
 However, this weak anisotropy of the inverse chiral structure can cause a problem at the same time, because the spin texture would easily turn into a multi-domain state, and the characteristic anomalous transports of non-collinear chiral antiferromagnets vanish due to mutual cancelation among contributions from different domains. \footnote{There is another non-collinear antiferromagnet, Mn${}_{3}$Ir where so far only anomalous spin Hall response~\cite{Zhange16007} is observed. We note that since the in-plane anisotropy in Mn${}_3$Ir is large, controlling the magnetic structure would be difficult~\cite{private}.}.
  
 In this letter, we discuss stability of the experimentally observed inverse chiral spin texture of non-collinear chiral antiferromagnets at finite temperature. We perform numerical calculations using stochastic Landau-Lifshitz-Gilbert (sLLG) equation~\cite{PhysRevB.58.14937,PhysRevB.91.134411} of spin dynamics for a simplified two-dimensional model of the spin structure in these materials. We find that the inverse chiral order is, once formed, highly stable against thermal fluctuations even without external magnetic fields, and its lifetime grows exponentially as we decrease the temperature. Moreover, we propose a way of controlling the in-plane magnetic structure with spin-current injection using the spin-filtering effect of ferromagnetic metals~\cite{Gijs:1997aa}. This offers a practical way of controlling the characteristic transport properties of non-collinear chiral antiferromagnets in a field-free way.
 \par
{\it Model and numerical method ---}
We consider a simplified model of non-collinear chiral antiferromagnets which describes planar spins in a single Kagom\'{e} layer obeying the following Hamiltonian~\cite{doi:10.1143/JPSJ.51.2478,Nagamiya:1982aa}:
\begin{align}
H &= \sum_{\langle i\alpha,j\beta \rangle } J (m^x_{i,\alpha}m^x_{j,\beta}+m^y_{i,\alpha}m^y_{j,\beta})- A\sum_{i\alpha}\cos(2\theta_{i\alpha}) \nonumber \\
&+D \sum_{\langle i\alpha,j\beta \rangle }\vec{d}_{i\alpha,j\beta}\cdot(\vec{m}_{i,\alpha}\times\vec{m}_{j,\beta}),
\label{Hamiltonian}
\end{align}
where $J > 0$ is the antiferromagnetic, nearest-neighbor exchange interaction, and $D$ is DM interaction.  We consider non-collinear states with in-plane spin-polarization, assuming large easy-plane anisotropy energy for each magnetic atom. According to the {\it ab-initio} calculation on Mn${}_3$Ir~\cite{PhysRevB.79.020403} the in-plane anisotropy energy is about $6$ meV, which is well larger than $D$ and $A$ which we use in this letter. We expect that even for Mn${}_3$Sn the in-plane anisotropy energy would not be so much different. Therefore, $m_{i,\alpha}^\mu$ represent $\mu = x, y$, and $z$ components of the classical spin with unit length in the site of $\alpha$-sublattice in the unit cell $i$, but $m_{i,\alpha}^z$ are always zero. The second term represents the in-plane anisotropy with strength $A$ in the direction of the nearest nonmagnetic atoms (e.g. Sn for Mn${}_{3}$Sn) of each site, and $\theta_{i\alpha}$ is the angle between $\vec{m}_{i,\alpha}$ and its easy-axis direction. 

As neutron scattering experiments show, when the external magnetic field is applied in the direction of the nearest non-magnetic atoms for spins in a particular sublattice, the ground state spin configuration becomes as shown in Fig.~\ref{spin config}($a$). That is, only one of three Mn spins in each unit cell points to the field direction and other two spins form the inverse chiral structure with small canting. This canting survives even without the external field due to the in-plane anisotropy and the system shows weak ferromagnetism in the single-domain inverse chiral state. This ground state configuration can be reproduced~\cite{doi:10.1143/JPSJ.51.2478} by choosing $\vec{d} = \hat{z}$ and when $D  >  A/\sqrt{3}$ is satisfied in the model Eq.~\eqref{Hamiltonian}. However, the stability of this inverse chiral structure in the absence of the external field at finite temperature, which is crucially important for device applications, remains unclear both theoretically and experimentally.

In this work, for the Hamiltonian~\eqref{Hamiltonian}, we numerically solve the following sLLG equation with Slonczewski-type spin-transfer-torque term~\cite{PhysRevB.58.14937,PhysRevB.91.134411,PhysRevLett.88.236601,PhysRevB.87.020402}:

%\begin{align}
%\frac{d\vec{M}_{i,\alpha}}{dt} &= - \gamma \vec{M}_{i,\alpha} \times \left(-\frac{\partial H}{\partial \vec{M}_{i,\alpha}} + \vec{h}_{T}\right)+ \frac{\alpha}{|\vec{M_{i,\alpha}}|}\vec{M}_{i,\alpha}\times \frac{d\vec{M}_{i,\alpha}}{dt} \nonumber \\+& \gamma B_{\mathrm{SL}} \left(\vec{M}_{i}\times \vec{p}\times\vec{M}_{i}\right).
%\label{LLGS}
%\end{align}

\begin{align}
\frac{d\vec{m}_{i,\alpha}}{d\tau} &= - \vec{m}_{i,\alpha} \times \left(-\frac{\partial \widetilde{H}}{\partial \vec{m}_{i,\alpha}} + \vec{h}_{T}\right)+ \alpha\vec{m}_{i,\alpha}\times \frac{d\vec{m}_{i,\alpha}}{d\tau} \nonumber \\+& \frac{j}{j_{0}}\left(\vec{m}_{i\alpha}\times \vec{p}\times\vec{m}_{i\alpha}\right),
\label{LLGS}
\end{align}
where $\widetilde{H} = H/J$ is the Hamiltonian Eq.~\eqref{Hamiltonian} normalized by the exchange coupling $J$. The time coordinate is normalized to be dimensionless: $\tau = tJ/\hbar$. The second term in the right hand side in Eq.~\eqref{LLGS} is Gilbert damping term describing dissipation with strength characterized by the dimensionless constant $\alpha_G$. In the framework of the sLLG equation, thermal fluctuations are treated as random fields $\vec{h}_{T}(t)$ satisfying
\begin{align}
\Braket{ h_{T}^{ \mu}(t)} &= 0, \nonumber \\
\Braket{ h_{T}^{\mu}(t)h_{T}^{\nu}(t')} &= \sigma \delta^{\mu, \nu}\delta(\vec{r}-\vec{r'})\delta(t-t'),
\label{random}
\end{align}
where $\mu, \nu = x, y, z$. Here $\sigma$ is determined by temperature from the fluctuation-dissipation theorem: $\sigma = 2 k_{B}T\alpha_G$/J. The numerical calculations are performed by Heum method with time step $\Delta \tau = 0.02$ using linearlization technique~\cite{Seki_BOOK}. For $ J = 15$ meV, $\Delta \tau$ corresponds to $0.02\hbar/J \sim 0.9$ fs.

The last term describes Slonczewski-type spin-transfer-torque from the injected spin-current (see for example, Ref.~\cite{PhysRevB.91.134411}). Here $j_{0} = \frac{2|e| M_{s} d J}{q \hbar^2 \gamma}$~\footnote{The exchange coupling $J$ enters here just because of the normalization of the time coordinate.} and $j$ is the density of the (spin-polarized) current injected along the $z$-direction with polarization $q$ and polarization direction $\vec{p}$. The gyromagnetic ratio is $\gamma$. We assume that the system is a thin film of thickness $d$ with saturation magnetization $M_s$, neglecting inter-layer interactions. This assumption does not exactly hold for known materials~\cite{refId0} but here we adopt that for simplicity. We expect that essential physics of the in-plane spin dynamics discussed in this letter is captured by the simplified model and the main results would be qualitatively the same as those for a more complicated, multi-layer model.

For example, if $M_s \sim 10^6 $A/m, $d = 1$ nm, $q=1$ (perfectly polarized), and $ J = 15$ meV, we have $j_{0}\sim10^{14}$ A$/\mathrm{m}^{2}$. In the following calculations, the system is a single-layer Kagom\'{e} lattice with 20 unit cells (60 sites in total), and periodic boundary condition is imposed.

{\it Stability of inverse chiral spin structure ---}
Here we study equilibrium properties of the present model in the absence of the spin-current ($j = 0)$. In the following, we measure $A$, $D$, and $T$ in the unit of $J$. That is, hereby $A$ means $A/J$ and so on. First, to determine numerical parameters we study the nature of weak ferromagnetism induced by the in-plane anisotropy. In Fig.~\ref{spin config}($b$), for several values of DM interaction, we show the numerically obtained weak ferromagnetic moment per site at zero temperature as a function of the anisotropy energy $A$. We assume that each Mn atom has magnetic moment $\simeq 3$ $\mu_{B}$. We see that the net magnetic moment depends linearly on the anisotropy energy. According to the neutron scattering experiment on Mn${}_3$Sn~\cite{doi:10.1143/JPSJ.51.2478},  we have $J = 15$ meV and $A = 0.01875$ meV and the ratio is $A/J = 0.0125$. From Fig.~\ref{spin config}, for $A/J = 0.00125$, we can reproduce the experimentally observed weak ferromagnetic moment ($\sim 2$ m$\mu_{B}$ per atom), by choosing  $D = 0.1 J = 1.5$ meV. 

  \begin{figure}[htbp]
   \centering
\includegraphics[width = 80mm]{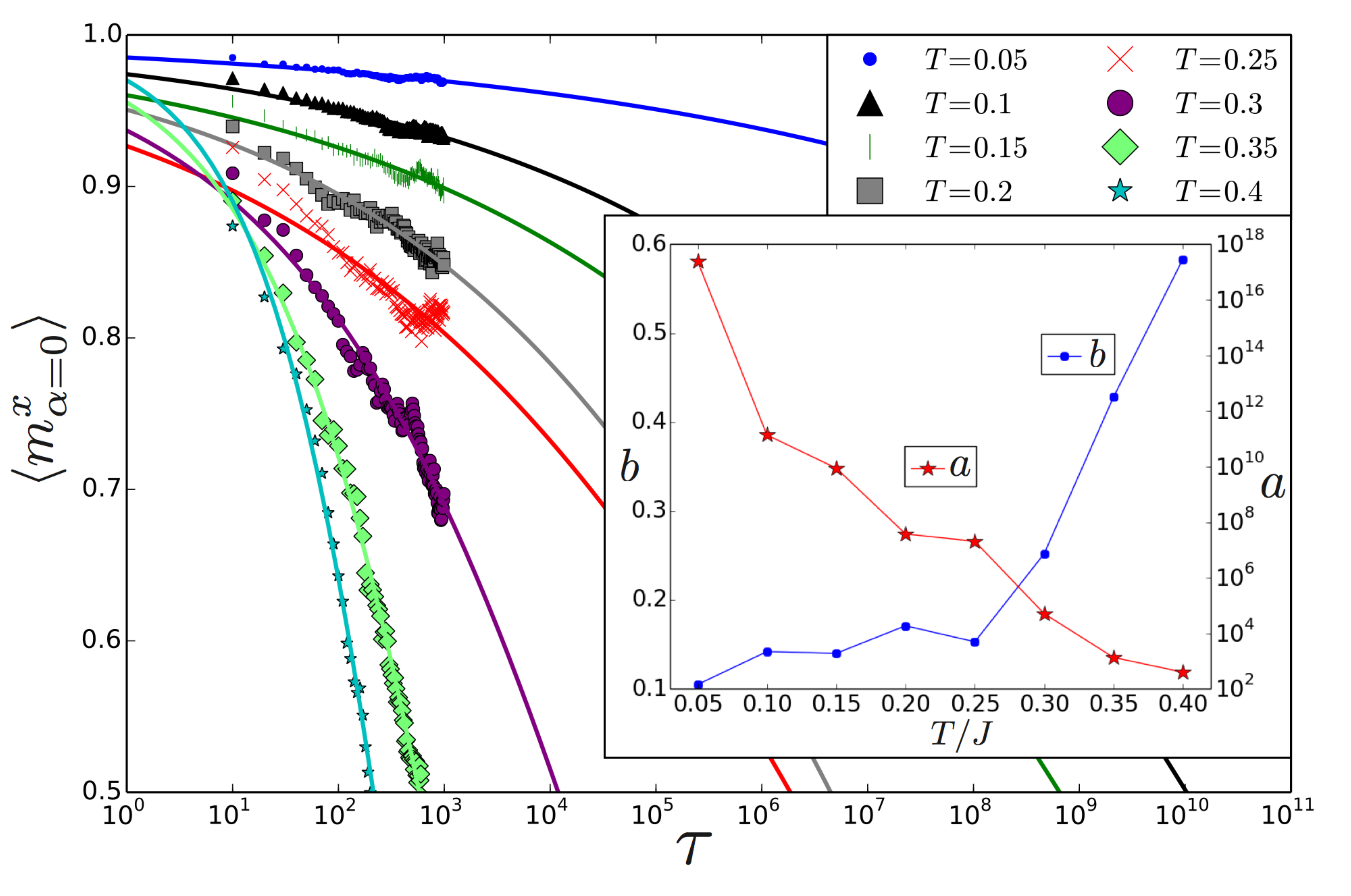}
       \caption{Stability of the inverse chiral order in the absence of the external field at finite temperatures for $D = 0.1$, $A = 0.00125$, and $\alpha_G = 0.1$. For each temperature we calculate the time evolution $10$ times and take the average (data represented by markers). Here $\langle m^x_{\alpha = 0} \rangle $ is the $x$-component of spins in the sublattice $\alpha = 0$ (see Fig.~\ref{spin config}), averaged over all the unit cells. The solid lines are the best fits for the numerical data with Eq.~\eqref{fit}. In the inset, the temperature dependence of fitting parameters is presented.}
          \label{combined} 
  \end{figure} 

Using the above parameters, next we consider the stability of this inverse chiral structure at finite temperature in the absence of the external field. Starting from the perfect inverse chiral state at $\tau = 0$, we calculate the time evolution until $\tau = 1000$ at finite temperatures without applying the external magnetic field. The planar spin model should work up to the temperature comparable with the easy-plane anisotropy. For $J=15$ meV, the anisotropy energy ($\sim 6$ meV for Mn${}_3$Ir) specifies the limit to be about $0.4 J$. To quantify how much the spin state at given time decay from the original state, we use $\langle m^x_{\alpha = 0} \rangle $, the average of $x$-component of spins in the sublattice $\alpha = 0$ [see Fig.~\ref{spin config}($a$)]. In the initial state, all spins in that sublattice point to the $x$-direction, so that $\langle m^x_{\alpha = 0} \rangle{}_{\tau = 0} = 1$.  For each temperature we calculate the time evolution $10$ times and take the average of them. The results are shown in Fig.~\ref{combined} using markers. We fit these averaged time evolutions assuming 
\begin{align}
\langle m^x_{\alpha = 0} \rangle{}_{\tau}  =\exp[- \left(\tau/a\right)^{b}]
\label{fit}
\end{align}
 with $a$ and $b$ being fitting parameters, from which we can estimate the lifetie of the initial state. We find that the lifetime grows exponentially as we decrease the temperature (see the inset of the figure). As the figure shows, even if the temperature is comparable with DM interaction, $T = 0.1$, the half-life $\tau_{0}$, defined by $\langle m^x_{\alpha = 0} \rangle_{\tau = \tau_{0}} = 0.5 $, would reach to $\tau_{0} \sim10^{10}$, which corresponds to $0.4$ ms for $J = 15$ meV. This extremely long lifetime allows us to perform various spintronic manipulations like spin-current injection which is discussed in the following. 

%As we argued, the easy-plane anisotropy energy is expected to be much larger than DM interaction and in-plane anisotropy energy. Therefore, the XY like effective spin model we are assuming can be used for discussing the relaxation process at finite temperature unless the temperature becomes high enough to compete with the anisotropy energy.

{\it Magnetization switching with spin-current ---}
    \begin{figure}[htbp]
   \centering
\includegraphics[width = 75mm]{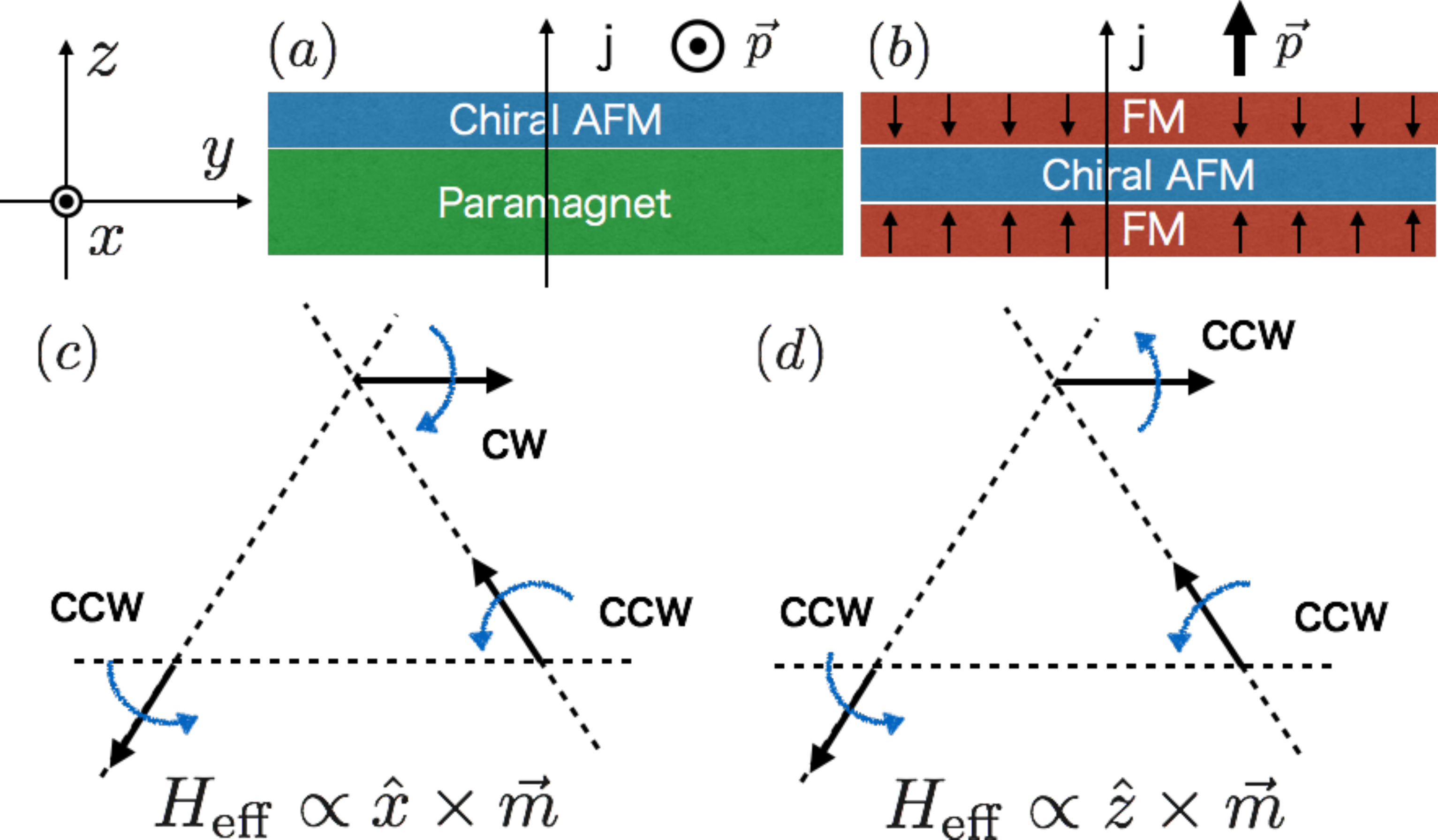}
       \caption{Two possible ways to inject spin-currents in chiral antiferromagnets. ($a$)($c$) Injection of pure spin-current with polarization $\vec{p}\perp\hat{z}$ using spin Hall effect in strongly spin-orbit-coupled paramagnetic metals. ($b$)($d$) Injection of spin-polarized current with $\vec{p}=\hat{z}$ using spin-filtering effect of ferromagnetic metals. For $\vec{p}\perp\vec{z}$, one of three spins in each unit cell is going to rotate in the opposite direction to others ($c$) while for $\vec{p}=\vec{z}$ all the spins rotate in the same direction ($d$).}
          \label{setup} 
  \end{figure}  
 We saw that the inverse chiral spin structure is highly stable even in the absence the external magnetic field. Given this observation, we study the possibility of controlling their magnetic structure without applying the magnetic field. For that purpose, we consider of injecting spin-current to the system which affects the spin texture through spin-transfer-torque.
 
Since the present system is antiferromagnetic and the spin structure is not spatially smooth, theoretical treatment of spin-transfer-torque in current-in-plane configuration: $\vec{j}\perp \hat{z}$ is subtle (see for example, Ref.~\cite{Zhang:2016aa}). Therefore, here we consider current-perpendicular-to-plane configuration: $\vec{j} \propto \hat{z}$. In this geometry, there are two ways to inject spin-current as shown in Fig.~\ref{setup}(a)(b). The first one is to use spin Hall effect~\cite{RevModPhys.87.1213} where we have pure spin current with polarization pointing inside the Kagom\'{e} plane: $\vec{p}\perp \hat{z}$ [Fig.~\ref{setup}(a)]. When $\vec{p} = (\cos\theta,\sin\theta,0)$, the spin-transfer torque term works as an effective magnetic field along the $z$-direction: $\vec{H}_{\mathrm{eff}} \propto (0, 0,m^y \sin\theta -m^x\cos\theta )$. Therefore, after we start injecting the spin-current, spins feel torque which is to rotate them around the $z$-axis in either clockwise (CW) or counter-clockwise (CCW) direction. However, as long as the effective field from the spin-current is weak compared to those from exchange and DM interaction, the inverse chiral spin structure of each unit cell is preserved and the net torque would be almost cancelled out [see Fig.~\ref{setup}(c)]. 

 The other setup is to inject spin-polarized electric current utilizing spin-filtering effect~\cite{Gijs:1997aa} of attached metallic ferromagnets [Fig.~\ref{setup}(b)]. Because of the different scattering rate between up and down spins in ferromagnets, electric current injected through the ferromagnetic layer becomes spin-polarized. In this case, the polarization is in the $z$-direction so that we have an in-plane effective field $\vec{H}_{\mathrm{eff}} \propto (-m^y,m^x,0)$. If we assume that the spins are well confined in the Kagom\'{e} plane, the precession around this effective field is suppressed. Hence, the effective field changes the in-plane direction of spins only through Gilbert damping towards the field-direction. In this case the direction of each spin and the effective field working on that are always orthogonal with each other: $\vec{m}\cdot \vec{H}_{\mathrm{eff}} = 0$, and three spins rotate in the same direction [see Fig.~\ref{setup}(d)], contrary to the former setup. Therefore, the spin-current will cause uniform rotation of the inverse chiral structure. We note that by using the synthetic antiferromagnetic configuration, Zeeman coupling with fields from ferromagnets, which is to break the in-plane nature of spins on each Kagom\'{e} layer, can be suppressed for a thin layer system [Fig.~\ref{setup}(b)]. Moreover, in this setup, we can achieve both CW and CCW rotation just by changing the sign of the applied voltage along the $z$-axis. 
 
\begin{figure}[htbp]
   \centering
\includegraphics[width = 80mm]{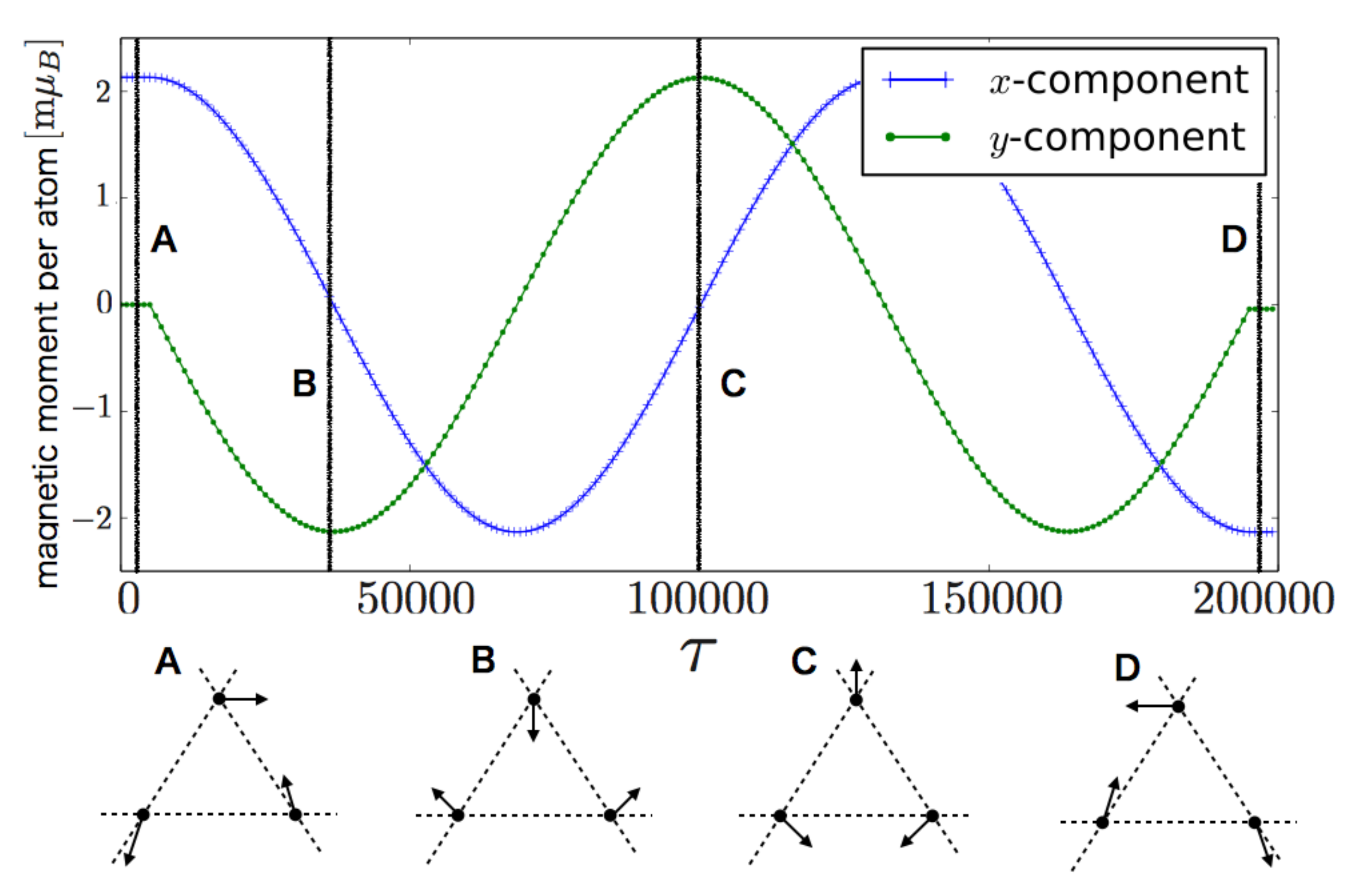}
       \caption{Time evolution of the magnetic moment per atom at zero temperature for $D = 0.1$, $A = 0.00125$, and $\alpha_G = 0.1$ in the absence of the external magnetic field. In the initial state at $\tau=0$, the system is weakly ferromagnetic in the $x$-direction ($\sim 2$ m$\mu_{B}$ per atom). We apply the static spin-current $j= 0.0005j_{0}$ with $\vec{p} = \hat{z}$ from $\tau=5000$ until $\tau = 195000$. For several points (A, B, C, and D) we schematically present snapshot spin structure of a unit cell.} 
          \label{rotation} 
  \end{figure}

Based on the argument above, we focus on the latter case with $\vec{p} = \hat{z}$ and see whether the spin-current can actually induce the rotation or not. Here we consider zero temperature and use fourth order Runge-Kutta method with time step $\Delta \tau = 0.2$ for the numerical calculation. In Fig.~\ref{rotation} we present the numerically obtained time-evolution of the net magnetic moment per atom at zero temperature for $D = 0.1$, $A = 0.00125$, $\alpha_G = 0.1$, and $j= 0.0005 j_{0}$, and the corresponding spin structure for several moments. Since the timescale of the rotation is much slower than other dynamics, at each time, the inverse chiral structures distort themselves to exploit the in-plane anisotropy energy, so that the net magnetic moment per atom changes accordingly. 

We see that the net magnetic moment and thus the in-plane magnetic structure rotate in the presence of the spin-current as we expected. Since the effective field is always perpendicular to the spin direction, to achieve deterministic switching we have to use a spin-current pulse with proper period. After turning off the current, the spin texture relaxes to one of six degenerate stable configurations determined by the in-plane anisotropy. Hence, we do not have to fine-tune the period and potentially six-fold magnetization switching is possible.

The threshold value of the current $j$ to induce uniform rotation is determined by the in-plane anisotropy energy. If we ignore all the inter-spin interactions, by considering the competition between the anisotropy and spin-current effect, the threshold value is given by $j_{\mathrm{th}} = j_{0} A/J$. For parameters assumed in this letter ($j_{0} \sim 10^{10}$ A$/$cm${}^2$ and $A/J = 0.00125$), we have $j_{\mathrm{th}} \sim 10^7$ A$/$cm${}^2$. Since the inverse chiral structure serves to significantly decrease the effective anisotropy of the inverse chiral structure~\cite{Nagamiya:1982aa,Nakatsuji:2015aa}, the actual threshold can be orders of magnitude smaller than this value. Indeed, the value $j = 0.0005j_{0}$ used in the calculation in Fig.~\ref{rotation} is much smaller than $j_{\mathrm{th}}$, but we see that spins can be rotated freely.
 
{\it Conclusion---}
We studied the spin structure of a simple two-dimensional model for non-collinear chiral antiferromagnets like Mn${}_{3}$Sn and Mn${}_{3}$Ge. On the basis of stochastic Landau-Lifshitz-Gilbert equation, we numerically calculated the time evolution of spins at finite temperatures. We found that the experimentally observed inverse chiral magnetic state of such systems is highly stable against thermal fluctuations even without the external magnetic field and its lifetime grows exponentially as we decrease the temperature. We also demonstrated that in the presence of spin-polarized current injected through attached metallic ferromagnets, the in-plane spin structure rotates almost freely. Since Kagom\'{e} chiral antiferromagnets are known to show gigantic anomalous Hall transports depending on the spin configuration inside Kagom\'{e} layers, our results offer a practical way of controlling transport properties of such materials in a field-free way.

{\it Acknowledgement---}
We thank S. Nakatsuji, S. Miwa, and M. Oshikawa for useful comments and discussions. H. F. is supported by Advanced Leading Graduate Course for Photon Science (ALPS) of Japan Society for the Promotion of Science (JSPS) and JSPS KAKENHI Grant-in-Aid for JSPS Fellows Grant No.~JP16J04752. This research was supported in part by the National Science Foundation under Grant No. NSF PHY-1125915.

\bibliography{ChiralAFM}
\end{document}